# The pseudogap: friend or foe of high $T_c$?

M. R. NORMAN*[1], D. PINES[2] and C. KALLIN[3]

[1]Materials Science Division, Argonne National Laboratory, Argonne, IL 60439, USA
[2]Institute for Complex Adaptive Matter, MS G754, Los Alamos National Laboratory, Los Alamos, NM 87574, USA
[3]Department of Physics & Astronomy, McMaster University, Hamilton, ON L8S 4M1, Canada

Abstract
Although nineteen years have passed since the discovery of high temperature cuprate superconductivity [1], there is still no consensus on its physical origin. This is in large part because of a lack of understanding of the state of matter out of which the superconductivity arises. In optimally and underdoped materials, this state exhibits a pseudogap at temperatures large compared to the superconducting transition temperature [2,3]. Although discovered only three years after the pioneering work of Bednorz and Müller, the physical origin of this pseudogap behavior and whether it constitutes a distinct phase of matter is still shrouded in mystery. In the summer of 2004, a band of physicists gathered for five weeks at the Aspen Center for Physics to discuss the pseudogap. In this perspective, we would like to summarize some of the results presented there and discuss the importance of the pseudogap phase in the context of strongly correlated electron systems.

Cuprates are remarkably simple materials. They are composed of $CuO_2$ planes separated by spacer layers. The low energy electronic structure of the planes is characterized by a single energy band [4]. But from this humble origin, a wealth of phenomena emerges (Fig. 1). In the parent material this energy band is half filled, corresponding to a $d^9$ configuration of the Cu ions (Fig. 2). Band theory would predict this to be a metal because of the odd number of electrons per Cu site, but the actual material is an insulator with a sizable energy gap of 2 eV. The origin of this insulating behavior was described many years ago by Nevill Mott [5] as a correlation effect. In the ground state, each Cu site would be in a $d^9$ configuration. Now, imagine one electron hops. In this case, the site left behind is in a $d^8$ configuration, whereas the site hopped to would have a $d^{10}$ configuration. The Coulomb energy cost for this process is known as the Hubbard U. If this Coulomb repulsion energy exceeds the kinetic energy gain due to hopping, the electrons become localized. The spins of these Cu ions form an antiferromagnetic arrangement (known as a Néel lattice) due to an interaction brought about by the virtual hopping of the antiparallel spins from one Cu ion to the next (the parallel configuration being disallowed by the Pauli exclusion principle). The energy gain due to this ordering is known as the superexchange energy, J, [6] which by the above argument can be seen to be proportional to $t^2/U$, where t is the hopping energy. The actual situation for the cuprates is somewhat more complicated, due to hybridization of the copper d and oxygen p orbitals. But the basic picture remains intact if we consider an "effective" Cu site that takes into account its surrounding oxygen neighbors (Fig. 2).

What happens when the cuprates are chemically doped? For the most studied "hole doped" case, $3^+$ cations in the spacer layers are replaced by $2^+$ ones. To maintain charge neutrality, electrons are pulled away from the $CuO_2$ planes, leaving behind holes. These holes can hop from one Cu site to the next, and by doing so, rapidly destroy the Néel lattice (each Cu spin removed from the plane breaks the magnetic bond with its four surrounding Cu spins). Then, remarkably, around 5% doping, a superconducting state emerges from this "destroyed" (or melted) antiferromagnetic state. The transition temperature to the superconducting state rapidly increases with further doping, reaching a maximum value (at about 15% doping) much greater than that of classic superconductors. Further doping causes the transition temperature to fall to zero (at about 25% doping), after which the material becomes a rather ordinary metal (known as a Fermi liquid).

The superconducting ground state is isomorphic with that of the theory of Bardeen, Cooper and Schrieffer [7], in that it consists of a condensate composed of Cooper pairs [8]. Photoemission experiments reveal sharp spectral peaks in the excitation spectrum [9], indicating the presence of quasiparticle-like states, which is also consistent with the long lifetime of electronic states as determined by various conductivity measurements [10,11]. Photoemission also sees the Bogoliubov-type dispersion of these states predicted by the BCS theory [12,13]. The one major difference is the symmetry of the condensate. In the original BCS theory, the pairs have an internal s-wave symmetry. However, the cuprate pairing condensate possesses d-wave symmetry, as conclusively demonstrated by several phase sensitive experiments [14,15]. Such higher angular momentum pairs have led to the speculation that the pairing in the cuprates has a different origin from that of classic superconductors. How do we get a handle on this question?

Superconductivity is an instability of the normal state. Therefore, to understand the origin of superconductivity, one must understand the nature of the normal state from which it arises. For the cuprates, this is where the real controversy begins to emerge. In 1989, three measurements of the spin response of cuprates at or below optimal doping were reported. In the NMR measurements by the Bell Laboratories group [2], a reduction was seen in the imaginary part of the low frequency dynamic spin susceptibility at a temperature large compared to the superconducting transition temperature. In the magnetic susceptibility measurements of Johnston [16] and the Knight shift measurements by the group of Henri Alloul [3], a similar suppression was seen in the uniform static susceptibility, albeit at significantly higher temperatures than those at which the low frequency dynamic spin response changed its character dramatically. Such depressions occur in classic superconductors where the pair state is a spin singlet, and thus the spin response is quenched as the temperature is lowered below $T_c$. The amazing thing about these experiments, though, is that this quenching phenomenon begins at a much higher temperature, $T^*$. The spin response shows no additional anomaly at $T_c$. That is, quenching analogous to spin singlet formation does not set in at $T_c$, but rather at $T^*$. Alternately, as Johnston first emphasized [16], the observed behavior follows that of a 2D Heisenberg antiferromagnet above its ordering temperature. The scaling behavior of the quenching of the static magnetic susceptibility predicted by the 2D Heisenberg model has been found for a broad spectrum of cuprates, including the undoped material [16-18] (Fig. 3). Over the years, measurements were made on a variety of different materials, and it was realized that $T^*$ increases with reduced doping, as opposed to $T_c$, which decreases with reduced doping. That is, quenching sets in at higher and higher temperatures as the Mott insulating phase is approached.

Thus, a central question emerged in the field. What is the relation of this "spin gap" (or pseudogap) phase to the superconducting phase? By continuity, one might expect that the pseudogap is also associated with spin singlet formation. But the opinions on this question have varied considerably. As any instability typically results in an energy gap, and such an energy gap by definition leads to a depression of the electronic density of states, some feel the pseudogap does not necessarily imply spin singlet formation. The onset of antiferromagnetic correlations would lead to similar behavior [16-18]. Even a charge density wave instability (or polaron formation for that matter) could lead to a pseudogap effect. It is amusing to note that one of the first theories of a pseudogap was for one dimensional charge density wave systems [19]. A direct probe of the density of states is provided by specific heat measurements, but the interpretation of these measurements is involved because of the necessity of subtracting off the much larger lattice contribution. They will be discussed later on in this article.

To delve further into these matters, experimentalists began to search for the pseudogap effect in probes which couple directly to the electronic charge, such as the electrical conductivity. Conductivity measurements can be made in two geometries, one which measures the planar charge response, the other the response for charge transport between the planes. Rather surprisingly, these two responses are quite different. Although the c-axis resistance shows an upturn for temperatures below T*, as would be expected from the presence of an energy gap, the planar resistivity actually drops [20].

More insight was obtained into this puzzling result by looking at the finite frequency response in the infrared range (Fig. 4). For the c-axis, the normal state response is roughly flat with frequency (implying incoherent behavior). As the pseudogap sets in, the conductivity at low frequencies gets suppressed, though the frequency at which the suppression sets in is roughly constant in temperature [21]. This leads to an effect which resembles the draining of water from a bathtub, and is quite novel, since for typical (second order) phase transitions, this frequency gap would vanish as T* is approached. This can be contrasted with the planar response. In this case, the Drude peak centered at zero frequency (signifying coherent charge transport in the plane) narrows (that is, becomes more coherent), leading to a dip in the conductivity between the Drude peak and the higher frequency response (sometimes referred to as the mid-infrared band) [22]. Although some advocate that this conductivity dip is the gap (with the narrower Drude peak below this gap representing "uncondensed" carriers), others use a generalized Drude analysis to argue that this "gap" should be more properly thought of as a gap in the scattering rate of the charged carriers [23]. This scattering rate gap leads to increased coherence of the low frequency carriers, and thus a narrower Drude peak.

This dichotomy between the planar and c-axis response has been taken as support of the "spin singlet" picture for the pseudogap phase [24]. The idea is that when the spins bind into singlets, the resulting spin gap reduces the scattering between the doped hole carriers, which leads to a narrower Drude peak in the pseudogap phase. On the other hand, spin and charge degrees of freedom must combine to form the actual electrons that tunnel between $CuO_2$ planes, thus a gap is seen directly in the c-axis response. This is consistent with angle resolved photoemission (ARPES) results, which also see an energy gap in the single-electron spectrum that "fills up" (rather than "closes") as T approaches T* [25].

In fact, ARPES reveals far more than this (Fig. 5). First, the energy gap is highly anisotropic around the Fermi surface, being large in the same regions of the Brillouin zone where the d-wave superconducting energy gap is large [26]. But unlike the superconducting energy gap, which vanishes at isolated points on the Fermi surface (the d-wave nodes), the pseudogap vanishes along segments of the Fermi surface centered at these nodes, known as Fermi arcs [25,27]. These disconnected arcs are quite difficult to understand, since any mean-field description would lead to zero energy contours that are continuous or closed in momentum space. This has led to the proposal that the pseudogap state is characterized by an unusual phase separation in momentum space, in which the "hot" electrons near $(\pi,0)$ are gapped below T*, and the "cold" electrons on the Fermi arc only become gapped below $T_c$. Since the c-axis response is primarily sensitive to the $(\pi,0)$ region (due to the in-plane momentum dependence of the c-axis coupling), this provides an alternate picture of the difference between the in-plane and c-axis optical responses. This "strong" pseudogap has a value comparable in size to the energy gap in the superconducting state (Fig. 5). In addition, though, ARPES sees a higher energy gap (known as the "hump" below $T_c$), sometimes referred to as the "weak" pseudogap [28]. The value of this gap, comparable in size to the superexchange J, adiabatically connects to the insulating gap seen by ARPES in the undoped phase [29]. This is strong evidence that the pseudogap is connected with Mott physics. Interestingly, this high energy gap is about four times the superconducting gap below $T_c$, and the two scale together with doping [28].

Moreover, as mentioned above, the superconducting state is characterized by the presence of sharp spectral peaks. These coherent peaks are observed to vanish above $T_c$ in the pseudogapped regions of the zone [25,26]. Taken at face value, the ARPES measurements would imply that the onset of superconductivity from the pseudogap phase is due to the onset of coherence. This would be consistent with recent infrared conductivity measurements, which indicate a boost in the in-plane optical spectral weight below $T_c$ for optimal and underdoped samples [30], though this result has been challenged by others [31]. The implied gain in the planar kinetic energy (as opposed to the decrease expected in BCS theory) is more than enough to account for the condensation energy of the superconducting state extracted from specific heat data.

All of the above experiments have been interpreted to be consistent with a so-called pre-formed pairs picture. In such a picture, the pairs would form at T*. The pairs would then condense (that is, become phase coherent) at the actual transition temperature, $T_c$. Note the analogy to a magnetic phase transition, where the moments form at a much higher temperature than the one at which they order. To investigate this picture, Phuan Ong's group [32] measured the Nernst effect, which is a higher order transport term representing the transverse electrical response to a thermal gradient in the presence of a magnetic field. In normal metals, the Nernst response is small, but in superconductors it is quite large due to the presence of vortices. Surprisingly, in the cuprates, the Nernst effect persists far above $T_c$ for underdoped compounds, indicating the presence of vortex-like excitations in the pseudogap phase. This vortex interpretation is further reinforced by new measurements by Ong's group of the diamagnetic response, which was found to track the Nernst signal [33]. However, the Nernst effect disappears well before T* is reached, and moreover, is rapidly quenched as $T_c$ falls with reduced doping, unlike T* which continues to rise (Fig. 6). This suggests that the Nernst effect is not associated with T*, but rather is a precursor to the superconductivity which arises within the pseudogap phase, in agreement with the results of terahertz measurements of the dynamic superfluid response by Joe Orenstein's group [34].

This brings us to the STM data. Scanning tunneling microscopy is similar to photoemission, in that it measures the c-axis current of single-particle-like excitations. Unlike ARPES, which is momentum resolved, STM is spatially resolved, and thus a complementary tool. The results have been equally spectacular. STM reveals the same pseudogap seen by ARPES [35], but also finds in a magnetic field the presence of the pseudogap below $T_c$ inside the cores of vortices (that is, the spectra exhibit a gap, but no coherence peaks) [36]. Further measurements, particularly in underdoped samples, have revealed the presence of strong spatial inhomogeneity, with "superconducting" regions of the sample (characterized by sharp spectral peaks), and "pseudogap" regions (characterized by a gap but no coherence peaks) that grow with reduced doping [37] (Fig. 7). The relation of this "phase separation in real space" to the "phase separation in momentum space" seen by ARPES is an intriguing issue. For instance, this electronic inhomogeneity is only observed for bias energies greater than $\Delta_{max}/2$, where $\Delta_{max}$ is the maximum of the anisotropic d-wave gap. Does this imply that the electronic states on the "Fermi arc" are homogeneous, forming local superconducting regions that are coupled through the pseudogapped regions?

More recently, STM studies have revealed the presence of weak charge modulations, which were first seen in vortices [38]. The modulation wavevectors (as revealed by Fourier transformation) are dispersive with energy in the superconducting state [39], but non-dispersive in the pseudogap state [40] (although one group claims a non-dispersive wavevector in the superconducting state as well [41]). The associated wavevectors are not commensurate with the lattice in the case of $Bi_2Sr_2CaCu_2O_{8+\delta}$ (Bi2212), but are in the case of $Na_xCa_{2-x}CuO_2Cl_2$ [42]. In the latter case, the modulations are more pronounced, and no coherence peaks have been observed yet (the samples being more heavily underdoped than the Bi2212 samples).

The observation of charge modulations has brought forth a host of questions. The modulation wavevectors are close to those predicted by the stripes model. In this model, the doped holes phase segregate into uni-directional stripes [43] separated by undoped antiferromagnetic regions, and this segregation has been advocated to explain a large body of experimental data in the cuprates [44]. A static stripe state does exist for one underdoped compound (Nd-doped $La_{2-x}Sr_xCuO_4$) [45], and stripes have been invoked to explain the dynamic spin incommensurability observed in inelastic neutron scattering data (Fig. 8), the incommensurability being due to the interruption of the periodic antiferromagnetic spin structure by the stripes. In this picture, the unusual "hourglass" shape of the dispersion of the spin excitations is taken to be a combination of spin waves associated with the stripe regions (lower part of the hourglass) and a triplon excitation associated with the antiferromagnetic regions which are modeled as a two leg spin ladder (upper part of the hourglass) [46]. Support for this picture is the similarity of observed spectra for non-superconducting [46] and superconducting [47] samples, though this hourglass dispersion can also be derived in uniform models that take into account the presence of a d-wave energy gap [48]. It is interesting to note that the incommensurate spin peaks seen by neutron scattering have the same dispersion with energy as the modulation wavevectors seen by STM [49]. On the other hand, the observed STM Fourier pattern appears more consistent with a two-dimensional checkerboard than uni-directional stripes [42], and this has been suggested as well for the neutron data [49]. With tongue in cheek, one could say that the cuprates have certainly earned their checks, but it remains to be seen whether they have earned their stripes [50]. However, it has been argued that, since the observed correlation length of this order is very short, the STM data are compatible with the stripes model in the presence of disorder where finite sized domains of the uni-directional stripes become pinned [44].

Even for a checkerboard modulation, many questions exist about its origin, especially in the $Na_xCa_{2-x}CuO_2Cl_2$ case where it is more pronounced. Does this represent a tendency towards Wigner crystallization of the doped holes? Or perhaps a Wigner crystal of pairs of holes? Scenarios that have been proposed to explain this include the pair density wave theories of Shou-cheng Zhang and co-workers [51], Zlatko Tesanovic [52], and Phil Anderson [53].

The new STM data have resurrected a debate that goes back to 1995 when the stripe state was first elucidated in Nd-doped LSCO [45]. Do charge effects drive spin effects, or vice versa? What we can say is that as the temperature is lowered, the pseudogap effect is first visible in experiments that either measure the spin response (NMR, inelastic neutron scattering), or the full electron response (ARPES, c-axis conductivity). Based on this, the general conclusion would be that for $T \sim T^*$, spin singlet formation, with its resulting spin gap, sets in. For lower temperatures, other effects begin to appear, such as charge modulations and the Nernst effect, the latter a precursor to the superconducting state that ultimately sets in at $T_c$. We note that although this seemed to be the consensus in Aspen, not all researchers (including some in Aspen) would agree with this. Chandra Varma [54] and Bob Laughlin [55] argue that one has a true phase transition to a new state of matter that is competitive with the superconducting state, while within the "spin singlet" picture, the various models differ in many important details. Moreover, an alternate phenomenology based on antiferromagnetic correlations has been proposed by Victor Barzykin and one of us [18, 56-58] to explain details of the susceptibility data. Their approach distinguishes two components. The first is a spin liquid that exists up to a doping level of about 20%; in the weak pseudogap regime between $\sim 2T^*$ and $\sim T^*/3$ it exhibits the scaling behavior shown in Fig. 3. At temperatures below $\sim T^*/3$, it enters the strong pseudogap regime characterized by a greater depression of the uniform susceptibility, and a depression of the NMR spin relaxation rate. In their scenario, this spin liquid competes successfully with superconductivity for the "hot" electrons near $(\pi,0)$. It is the remaining electrons that go superconducting; these are responsible for the doping dependent condensate fraction, and the temperature independent contribution to the uniform susceptibility seen by Johnston [16] and Nakano [17].

We illustrate this debate in Fig. 9, which represents various theoretical idealizations of the observed phase diagram. In the three examples, the T* "phase" line either (1) is degenerate with $T_c$ on the overdoped side, (2) cuts through the $T_c$ dome (eventually vanishing inside the dome at a T=0 "quantum critical point"), or (3) ends at the $T_c$ dome. When considering these diagrams, it should be noted that in most models, the T* line is a crossover line, rather than a true phase line. As such, T* can become ill-defined at high temperatures (where the crossover is broadened by thermal effects) or at low temperatures (where superconducting fluctuations set in).

Diagram (1) was the original idea of the spin singlet scenario [59], and also has been advocated in those scenarios that assert that the pseudogap phase is a precursor to the superconducting state, that is pairing without long range phase coherence [60-61]. In these scenarios, illustrated in the left panel of Fig. 10, there are two crossover lines: T* where spin singlets (or pairs) form, and $T_{coh}$, where the doped holes (or pairs) become phase coherent. Only below both crossover lines is superconductivity possible (due to spin-charge recombination in the spin singlet model, or due to phase coherence of the pairs in the precursor superconductivity model). An important difference, though, is that in the spin singlet scenario, the pseudogap phase is considered to be a "mother" phase out of which either the antiferromagnetic phase arises at low dopings, or the superconducting phase at higher dopings [24]. Similar considerations occur for the SO(5) model of Shou-cheng Zhang and collaborators [62] which attempts to unify antiferromagnetism and superconductivity.

An important aspect of the above scenarios is the general doping dependence of the various phase lines. The energy gain associated with spin singlet formation is the superexchange energy, J. This energy competes with the kinetic energy of the doped holes, ~tx, where t is the hopping energy of the hole and x the number of doped holes per Cu ion. This would imply that the T* line is proportional to J-tx, in agreement with Fig. 1 [63]. As the phase operator is conjugate to the number operator, this implies that $T_{coh}$ would be proportional to x [64]. This explains the suppression of $T_c$ on the underdoped side of the phase diagram.

Diagram (2) is what one would expect in scenarios in which another state of matter competes with superconductivity; i.e. the pseudogap and the superconducting gap compete for the same Fermi surface, so as one wins, the other loses. In the antiferromagnetic scenario, the pseudogap phase reflects the onset of strong antiferromagnetic correlations, but without long range order [58,65] while in the spin singlet scenario it is a liquid of spins without long range order (the original RVB idea of Anderson [66]). In some competitive scenarios, T* would be a true phase line (the d-density wave state proposed by Sudip Chakravarty and co-workers [67], or the orbital current state advocated by Chandra Varma [68]). In the latter scenarios, it is difficult for conventional probes to couple directly to the order parameter, thus the term "hidden order", a concept which has also been used in the context of certain heavy fermion materials [69] that have similar phase diagrams. These scenarios have a $T_{coh}$ line as well (right panel of Fig. 10) separating the "quantum disordered" (i.e., Fermi liquid) phase from the "quantum critical" (i.e., normal state) phase [70]. Of particular relevance for the cuprates is the observation of linear T resistivity over a wide temperature range in the normal state [71], along with linewidths in photoemission spectra that grow linearly with both temperature and binding energy [72], both of which are classic signatures of "quantum critical" scaling [73]. In that vein, it is of interest to note evidence of quantum critical scaling as well in infrared conductivity [74] and inelastic neutron scattering data [75].

What is the experimental evidence for long range order in the pseudogap phase? There have been some reports of an ordered moment from elastic neutron scattering, although the moment value is much smaller than in the Néel state [76]. On the other hand, the results are quite sensitive to sample quality, indicating the possibility that this is a disorder effect (neutrons require large single crystals, which are difficult to make homogeneous). Circularly polarized ARPES has also revealed the possibility of time reversal symmetry breaking associated with T* [77], but the results have been challenged by others [78]. The orbital current pattern consistent with such ARPES measurements has recently been claimed to have been seen directly in neutron scattering measurements [79].

Specific heat data have frequently been cited in support of diagram (2) [80]. Specific heat has the advantage of being a bulk probe of the electronic degrees of freedom, but as it is dominated by the lattice contribution, a large subtraction is necessary to identify the electronic component. The subtracted data appear to represent a (states non-conserving) suppression of the density of states in the pseudogap phase [80]. T* is interpreted as the energy scale for the pseudogap, rather than a temperature at which the pseudogap disappears. Moreover, Jeff Tallon and John Loram have advocated that the T* line actually cuts through the $T_c$ dome [81], citing experiments that show the $T_c$ dome collapsing about the T* line as impurities are introduced (Fig. 11), but the existence of a T* line below $T_c$ is certainly controversial. In particular, they believe T* terminates in doping at a quantum critical point, and provide various thermodynamic evidence for this, such as a peak in the superconducting condensation energy extracted from the specific heat data (Fig.11). Two possible indicators for a distinct T* line below $T_c$ are the appearance of antiferromagnetic order upon applying a magnetic field in underdoped samples [82], and the observation of strong electronic inhomogeneity by STM for underdoped samples [37], but whether these behaviors correlate in temperature and doping with the purported T* line below $T_c$ has yet to be addressed. Above $T_c$, specific heat data indicate a loss of entropy at T* [80], but nothing like a phase transition anomaly (though some specific heat measurements show a weak bump at T* [83]).

One signature of scenarios in which the pseudogap competes with superconductivity for the same electrons is the appearance of two energy gaps below $T_c$ in underdoped samples, a superconducting gap and a pseudogap. From ARPES, though, there is no evidence for this at the present time. Below $T_c$, one sees a simple d-wave gap, with no additional momentum or energy structure that would indicate the presence of two energy gaps (Fig. 5). From these data, one comes to the conclusion that either the pseudogap turns into the superconducting gap below $T_c$, or is replaced by the latter. It has been difficult to distinguish these two possibilities because the two energy gaps are comparable in size, and in some sense one is doing an apples and oranges comparison. That is, above $T_c$, there are no coherent features, and the pseudogap is simply the leading edge of the broad incoherent spectrum; below $T_c$, coherent peaks form and the superconducting gap is the location of the spectral peak. Does this imply these two gaps are separate phenomena? Certainly, there are differences in their angular anisotropies (Fig. 5).

This brings us to diagram (3), a possibility that has not been addressed much in the literature. In this case, T* would terminate at the $T_c$ dome (or "peter out" before reaching it), and the pseudogap would be replaced by the superconducting gap below $T_c$. To test this would require careful data sampling as a function of doping to look for the implied "kink" in the phase diagram as one passes from the T* line to the $T_c$ line near optimal doping.

So, what experiments could be done to address these various issues? In spin singlet scenarios in which the pseudogap represents a spin liquid phase, there is the possibility of various topological excitations that would characterize such a state. Todadri Senthil and Patrick Lee have proposed an experiment to look for the resulting "gauge" flux associated with these excitations [84]; an earlier search for signatures of topological excitations ("visons") was unsuccessful [85]. If the pseudogap phase represents some sort of hidden order, then it should prove possible to identify it experimentally. Orbital currents, for instance, would give rise to weak magnetic fields that are in principle detectable and distinguishable from the spin moments on the Cu sites. Perhaps the "bump" seen in certain specific heat measurements [83] would become more pronounced if the samples were more homogeneous. The group at UBC, for instance, has been able to grow high quality $YBa_2Cu_3O_{6+x}$ samples well into the underdoped regime, but up to now, these have been primarily used to study phenomena at lower temperatures [86]. The data on these samples have revealed the possibility that some analogue of the nodal d-wave superconducting state persists for dopings less than that where superconductivity disappears [87]. Does this mean the low temperature pseudogap state is a "nodal metal"? Certainly, it would be very interesting to see what such samples would reveal for temperatures comparable to T*.

Improvements in various spectroscopic measurements also give the possibility of resolving the pseudogap debate. Inelastic neutron scattering, which measures the spin response of the system as a function of momentum and energy, also sees the spin gap phenomena, as well as a "spin resonance" feature that, though strongest below $T_c$, has a precursor signature which develops at T* [88]. Associated with this commensurate spin resonance are dispersive incommensurate excitations at energies both below and above the spin resonance energy as shown in Fig. 8 [46-49]. A detailed study of these various excitations between $T_c$ and T*, and their relation to the charge modulations seen by STM, should prove to be most revealing.

In ARPES, the advent of high momentum and energy resolution detectors has led to a wealth of new data [9]. Of particular interest would be a detailed mapping of both the superconducting and pseudogaps as a function of temperature, momentum, and doping at high resolution, which would tell us a lot more about the nature of these two energy gaps and how they are related. Moreover, a better understanding of the nature of the "Fermi arcs" would be most welcome.

Also of interest would be a robust handle on the purported $T_{coh}$ line. So far, the only evidence has been from planar resistivity data that find a deviation from the linear T relation for $T < T_{coh}$ [89], and ARPES where coherent features first appear in the spectra [90]. No reports from other probes have been reported, and the doping dependence of this line is ill defined at present.

We now turn to a series of open questions for both experiment and theory to consider.

1. First, and perhaps most important, is the pseudogap a friend or a foe of high $T_c$? If foe, can a way be found to eliminate it and thereby achieve even higher superconducting transition temperatures?

2. Is there long range order associated with T*? Can one define an "order parameter" for the pseudogap phase so that a mean field description of the pseudogap phase is possible?

3. Is the pseudogap phase a spin liquid with topological excitations or, more generally, a non-Fermi liquid?

4. If T* cuts through the $T_c$ dome as a function of doping, is the superconducting state different on opposite sides of the T* line?

5. Is charge ordering an epiphenomenon, or is it central to the pseudogap phenomenon?

6. Is the undoped phase really a Mott insulator? Are the states observed inside the insulating gap merely a disorder effect, or of a more intrinsic nature?

7. What is the role of the lattice? Can phonons be as easily dismissed as many theorists assume? Are polaronic effects involved?

In regards to this last point, recent ARPES experiments on underdoped and undoped samples have led Z. X. Shen's group to advocate that the "high energy" pseudogap is not a precursor to the Mott gap, but rather a polaronic shift of the electronic states due to strong electron-lattice coupling [91].

Finally, how do we combine theory with experiment to obtain a full understanding of the pseudogap? Obviously, new experimental data will play a major role here, but the general feeling is that progress will only be made if both a consistent phenomenology can be developed and theories become more rigorous. We are, in a sense, not as far advanced as the superconducting community was at the time of John Bardeen's Encyclopedia of Physics article [92], in which he showed that with an energy gap in the spectrum, most of the experimental data could be understood phenomenologically.

Without a consistent phenomenology and detailed calculations to compare to experiment, a given model is more an idea than a theory. Of course, rigorous calculations can be difficult, as many proposed theories are non-perturbative in nature. Such is the challenge of strongly correlated electron systems, where the ground states often do not appear to be simple Landau Fermi liquids. The development of classic many-body theory in the 1950s together with a consistent phenomenology led to the BCS solution of 1957. Similar developments in the theory of strongly correlated electron systems should lead to a solution of the high $T_c$ problem. The authors of this perspective believe the path to that solution will involve an explanation of the pseudogap phenomenon, and that "spin" physics will play a major role in that solution. Only time will tell whether we are correct or not. Regardless, it will be a lot of fun getting there [93].

**Acknowledgments:** This perspective was a result of a workshop at the Aspen Center for Physics, whose support the authors gratefully acknowledge. We would also like to acknowledge support from the US Department of Energy, Office of Science, under Contract No. W-31-109-ENG-38, the Institute for Complex Adaptive Matter, and the Natural Sciences and Engineering Research Council of Canada.

*Correspondence should be addressed to M.R.N. (e-mail: norman@anl.gov).

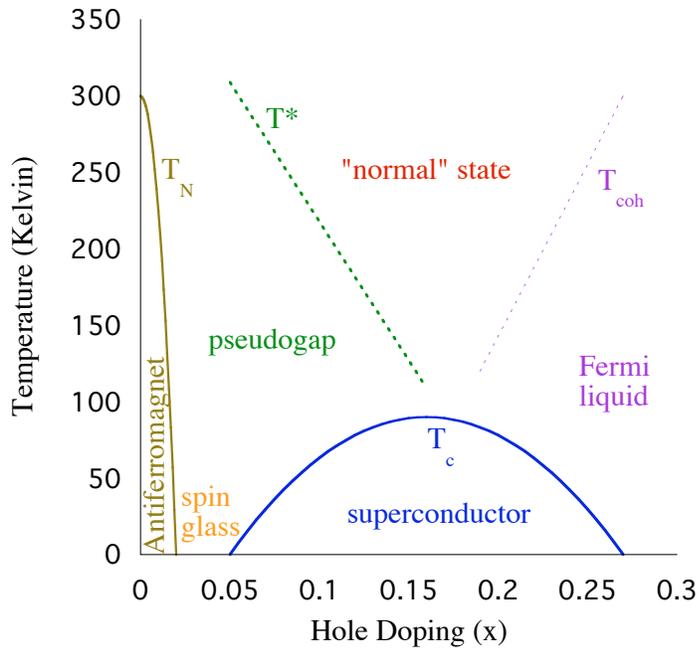

Figure 1 – Phase diagram of the cuprates versus x, the number of doped holes per Cu ion. Solid lines represent true thermodynamic phase transitions; dotted lines indicate crossover behavior.

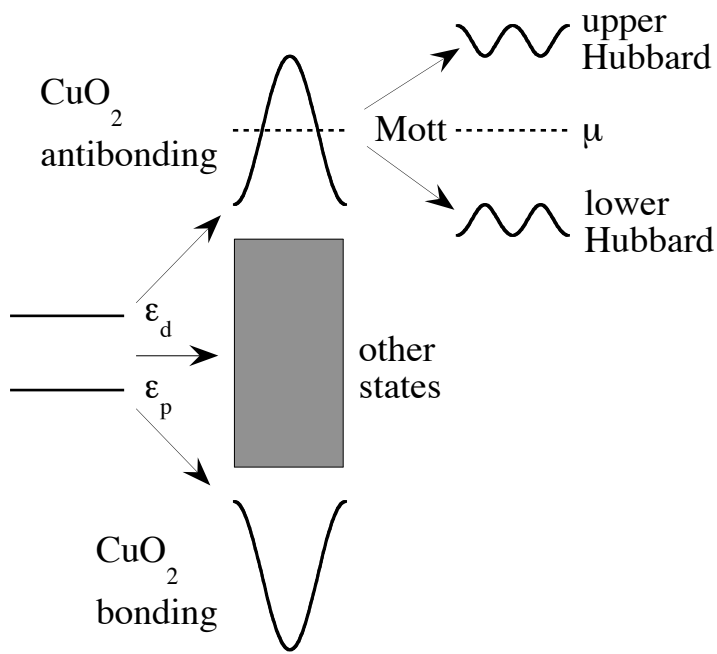

Figure 2 – Electronic structure of the cuprates. The copper d and oxygen p levels hybridize, resulting in a half filled antibonding band. A Mott gap due to Coulomb correlations splits this band, leading to the formation of an upper Hubbard band and a lower Hubbard band, with the chemical potential, $\mu$, inside this gap in the undoped material. Adapted from Ref. 48.

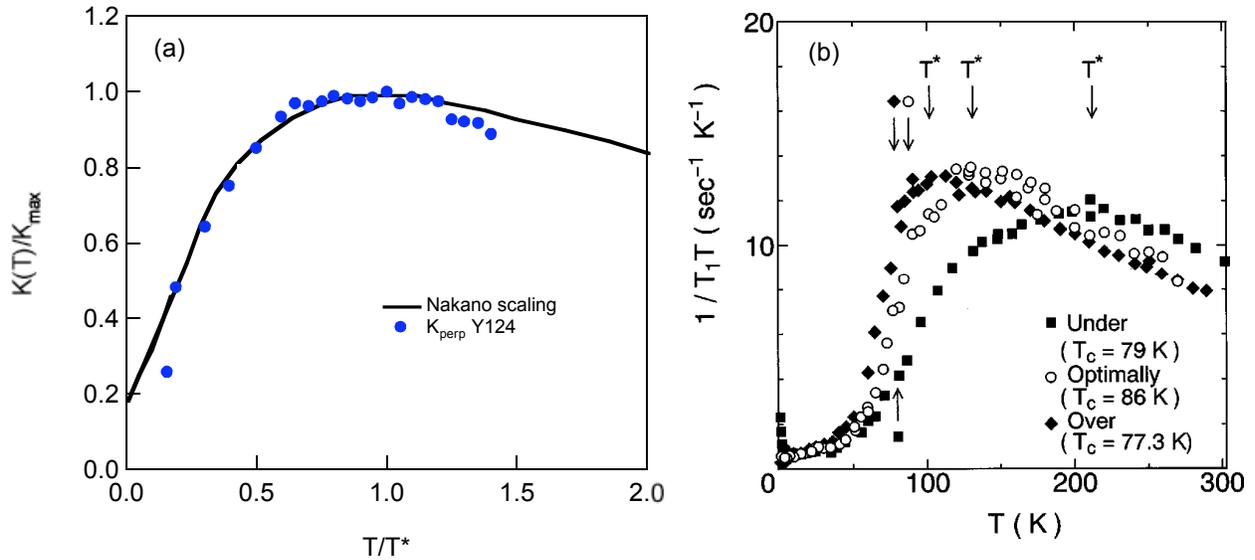

Figure 3 – (a) Copper Knight shift data for $YBa_2Cu_4O_8$ [94] compared to the scaling curve of Johnston [16] and Nakano *et al.* [17] based on $La_{2-x}Sr_xCuO_4$ data. (b) NMR spin relaxation rate for Bi2212 as a function of doping [95]. Note the suppression of the spin response below T*, and the smooth behavior through $T_c$.

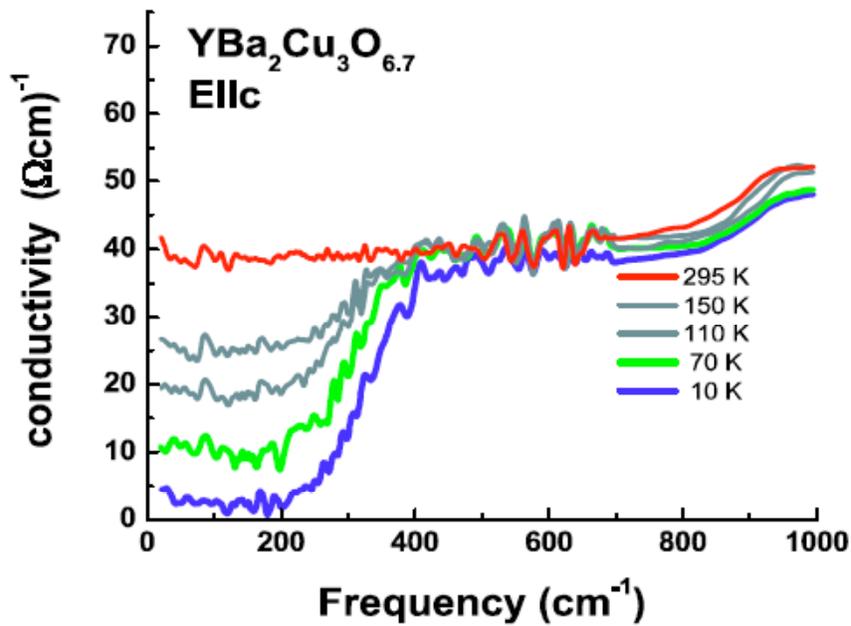

Figure 4 – c-axis conductivity (phonon subtracted) for underdoped YBCO, with a pseudogap that fills in with increasing temperature [21]. Figure courtesy of Dmitri Basov.

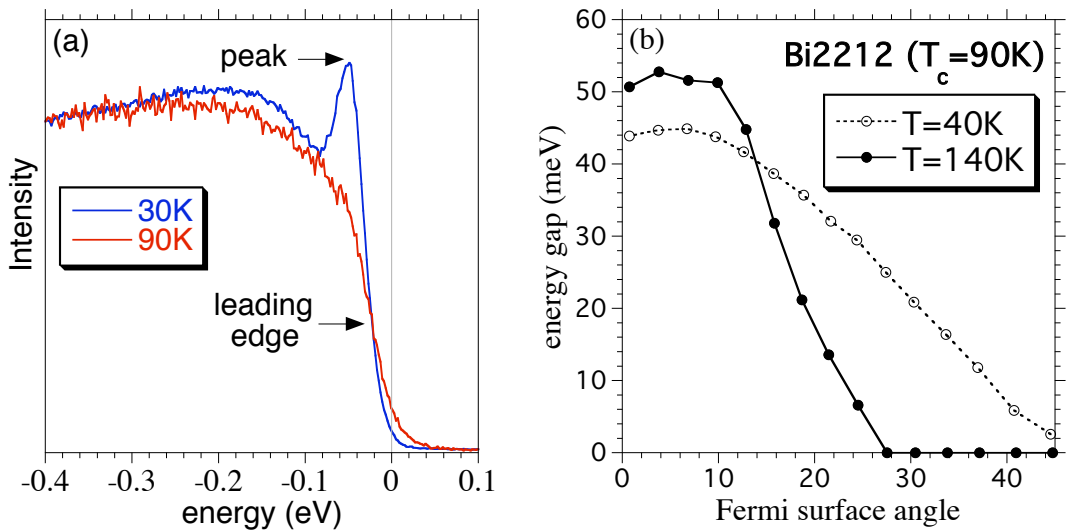

Figure 5 – (a) ARPES spectrum at (π,0) for an underdoped Bi2212 sample in the superconducting state (30K) and the pseudogap phase (90K). The sharp peak in the superconducting state is replaced by a leading edge gap in the pseudogap phase. (b) Angular anisotropy of the superconducting gap (40K) and the pseudogap (140K) for an optimal doped Bi2212 sample. Data courtesy of Adam Kaminski and Juan Carlos Campuzano.

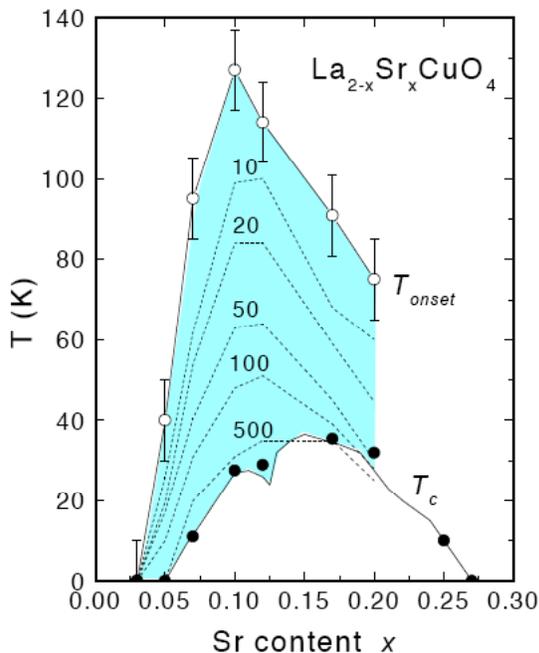

Figure 6 – Contours of the vortex-like Nernst signal for LSCO versus hole doping. Note that unlike T*, the Nernst effect has roughly the same trend with doping as $T_c$ does [96].

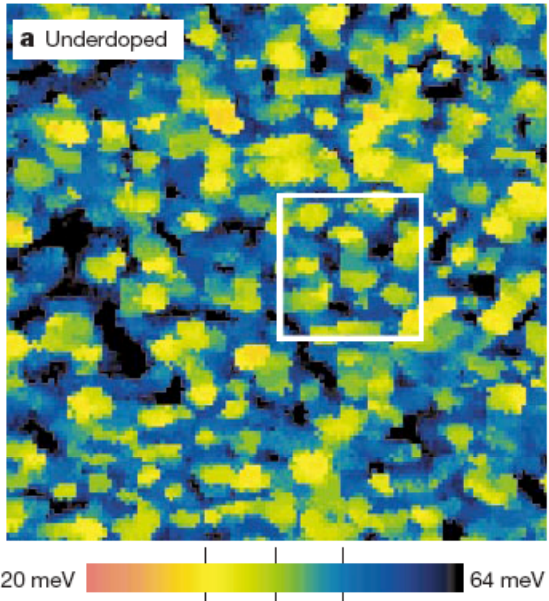

Figure 7 – Variation of the energy gap in scanning tunneling microscopy versus spatial position in an underdoped Bi2212 sample [37]. The high gap regions exhibit pseudogap-type spectra, the lower gap regions exhibit spectra more typical of a coherent superconductor.

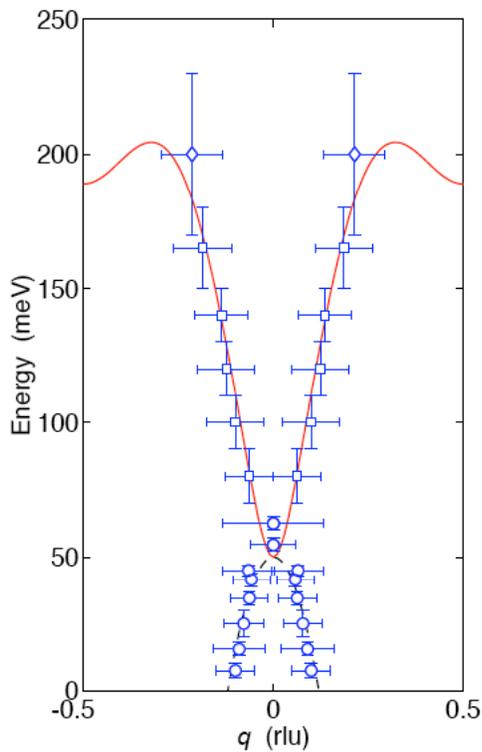

Figure 8 - Hourglass like dispersion of the spin excitations in the stripe ordered phase of $La_{1.875}Ba_{0.125}CuO_4$, measured with respect to $Q=(\pi,\pi)$, as revealed by inelastic neutron scattering [46]. Similar dispersions are seen in the superconducting state of YBCO [47].

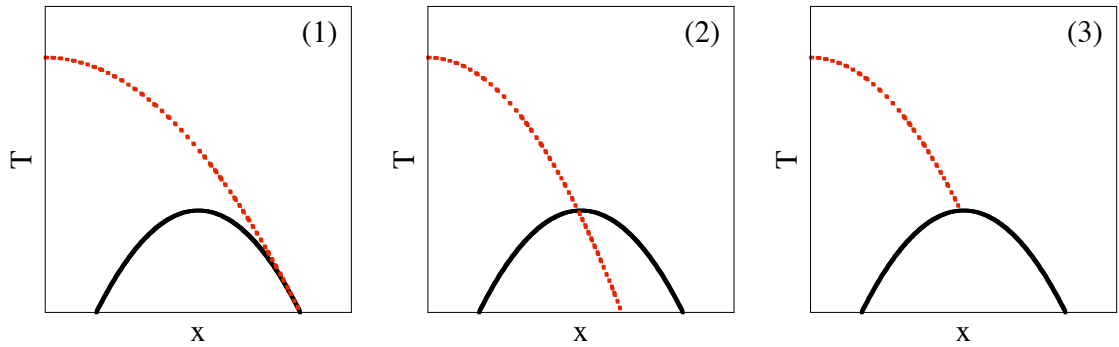

Figure 9 – Three possibilities for the phase diagram of the cuprates. The solid black line is the superconducting transition temperature, $T_c$, and the red dashed line the pseudogap phase line, $T^*$. In most scenarios, the $T^*$ line should be taken as a crossover, rather than a true phase transition.

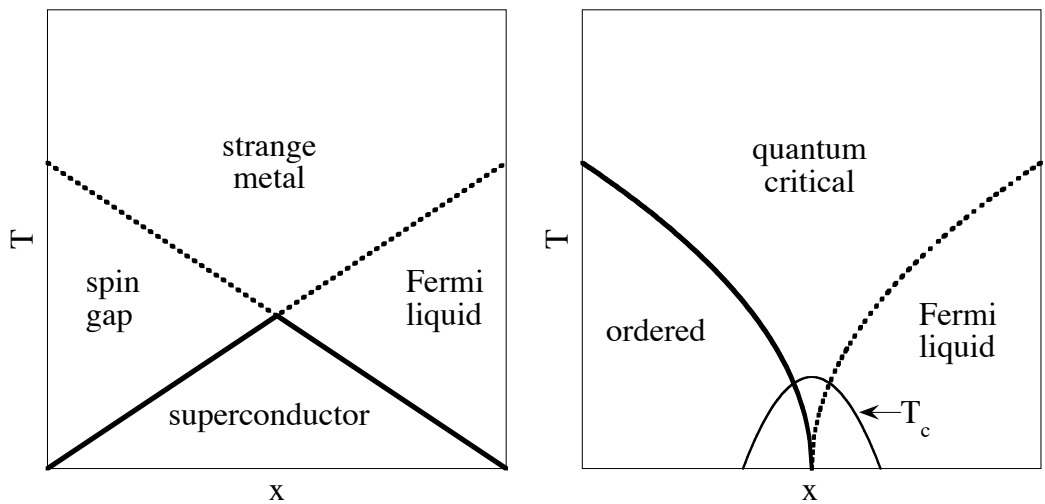

Figure 10 – Two proposed theoretical phase diagrams for the cuprates. RVB picture (left panel) and quantum critical scenario (right panel) [48].

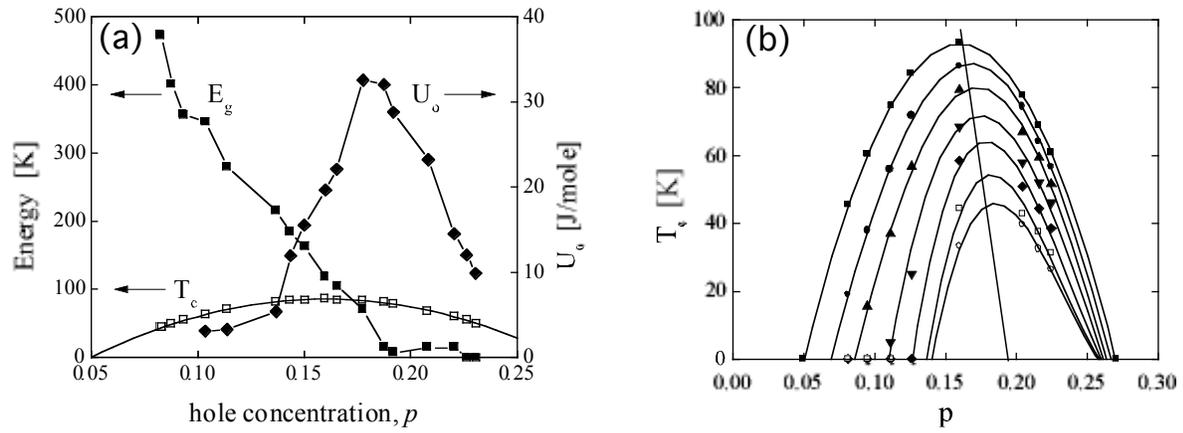

Figure 11 – (a) the pseudogap energy scale, $E_g$, and the superconducting condensation energy, $U_0$, as derived from specific heat data in YBCO, versus hole doping [81]. (b) Collapse of the superconducting dome about the $E_g$ crossover line with increasing cobalt impurities for Bi2212 [81].